\documentclass{webofc}
\usepackage[varg]{txfonts}   
\usepackage{xspace}
\usepackage[colorlinks]{hyperref}
\usepackage{graphicx}
\usepackage{relsize}
\usepackage{subcaption}

\newcommand{\linkk}[2]{\href{#1}{#2}\xspace}
\newcommand{\Rplus}{\protect\hspace{-.1em}\protect\raisebox{.35ex}{\smaller{\smaller\textbf{+}}}}
\newcommand{\cpp}{\mbox{C\Rplus\Rplus}\xspace}

\graphicspath{{../../../common_resources/figures}}

\begin{document}
\title{Towards podio v1.0 - A first stable release of the EDM toolkit}

\author{\firstname{Juan Miguel} \lastname{Carceller}\inst{1}
        \and \firstname{Frank} \lastname{Gaede}\inst{2}
        \and \firstname{Gerardo} \lastname{Ganis}\inst{1}
        \and \firstname{Benedikt} \lastname{Hegner}\inst{1}
        \and \firstname{Clement} \lastname{Helsens}\inst{1,3}
        \and \firstname{Thomas} \lastname{Madlener}\inst{2}\fnsep\thanks{\email{thomas.madlener@desy.de}}
        \and \firstname{Andr\'e} \lastname{Sailer}\inst{1}
        \and \firstname{Graeme A} \lastname{Stewart}\inst{1}
        \and \firstname{Valentin} \lastname{Volkl}\inst{1}
}

\institute{CERN, Switzerland
  \and Deutsches Elektronen-Synchrotron, Germany
  \and KIT, Germany
}

\abstract{%
  A performant and easy-to-use event data model (EDM) is a key component of any
  HEP software stack. The podio EDM toolkit provides a user friendly way of
  generating such a performant implementation in C++ from a high level
  description in yaml format. Finalizing a few important developments, we
  are in the final stretches for release v1.0 of podio, a stable release with backward compatibility for
  datafiles written with podio from then on. We present an overview of the podio
  basics, and go into slighty more technical detail on the most important topics
  and developments. These include: schema evolution for generated EDMs,
  multithreading with podio generated EDMs, the implementation of them as well
  as the basics of I/O. Using EDM4hep, the common and shared EDM of the Key4hep
  project, we highlight a few of the smaller features in action as well as some
  lessons learned during the development of EDM4hep and podio. Finally, we show
  how podio has been integrated into the Gaudi based event processing framework
  that is used by Key4hep, before we conclude with a brief outlook on potential
  developments after v1.0.}
\maketitle

\section{Introduction}\label{sec:intro}

A typical high energy physics (HEP) analysis workflow comprises many steps,
where each is usually in charge of producing a specific result or of linking
these results into a more comprehensive picture of the underlying physics event.
A core part of such a software stack is the so called event data model (EDM)
that enables the flow of information between the different components and also
defines the language of this communication. Crucially, it is also the language
in which users express their ideas. Defining the contents of such an EDM is one
aspect, the other one is the efficient implementation of the schema that users
come up with. The podio EDM toolkit~\cite{podio_2020,podio,podio:github} addresses the latter
part, while EDM4hep~\cite{Gaede:2021izq,Gaede:2022leb} defines the common EDM of
the Key4hep project~\cite{FernandezDeclara:2022voh,key4hep_2021} using podio.

In these proceedings we focus on recent developments in podio towards a first
stable release, which we plan to announce in the near future after finishing the
necessary developments. We will discuss these final missing pieces in the text
where appropriate.

The structure of these proceedings is as follows: After a brief recap of the
basics of podio in Section~\ref{sec:podio_basics} we describe the Frame concept
and how it relates to the multithreading concept of podio in
Section~\ref{sec:frame}. In Section~\ref{sec:schema_io} we dive into some of the
challenges and solutions we found in the context of implementing a schema
evolution mechanism for EDMs generated by podio, before we briefly talk about
other recent developments in Section~\ref{sec:recent_devs}. These include the
addition of a new RNTuple~\cite{Blomer:2020usr,Lopez-Gomez:2022umr} based I/O
backend, as well as the integration of Frame based I/O into the core framework
of Key4hep~\cite{zenodo:k4fwcore}. We close with a discussion of the open issues
that we still want to address before actually releasing version 1.0 of podio.

\section{The basics of podio}\label{sec:podio_basics}

Since the basic ideas of podio as well as the features of the generated code
have been covered in more detail in previous
publications~\cite{podio,Gaede:2021izq,podio_2020}, we will focus on those parts
that are most relevant for the technical discussions later.

\begin{figure}[htbp]
  \centering

  \subcaptionbox{podio layers
    \label{fig:podio-layers}}{\includegraphics[width=0.4\linewidth]{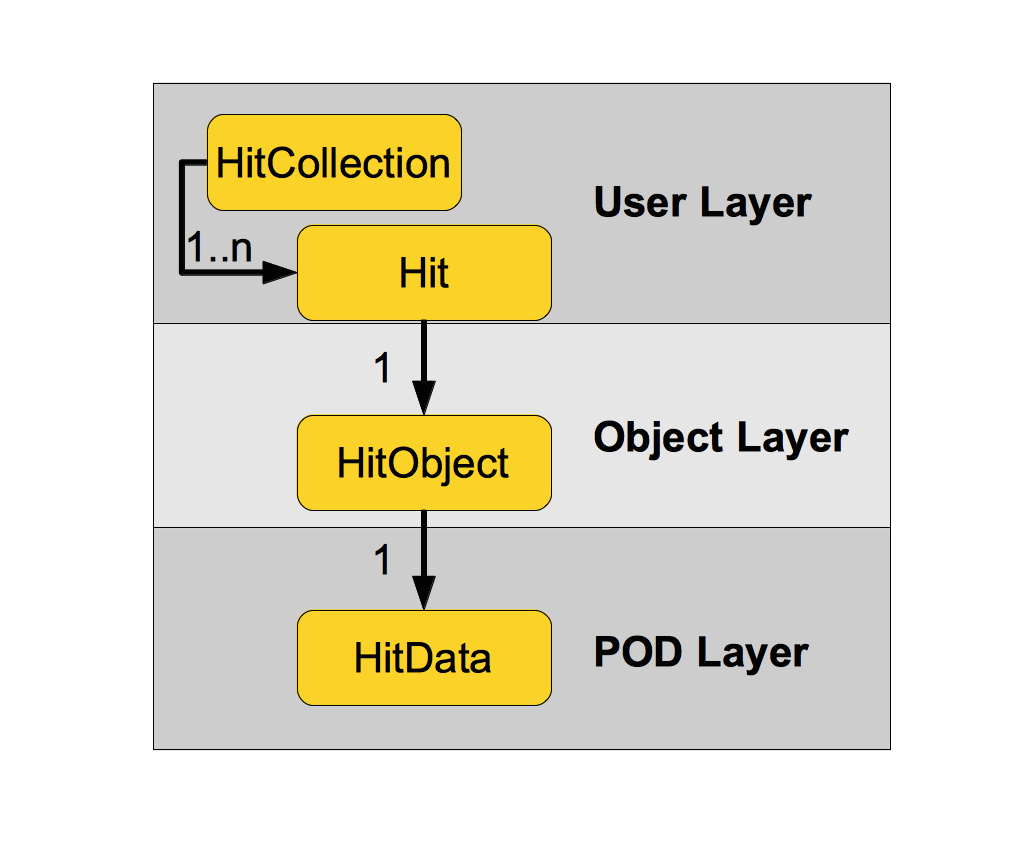}}
  \subcaptionbox{schematic collection I/O
    \label{fig:collection-io}}{\includegraphics[width=0.4\linewidth]{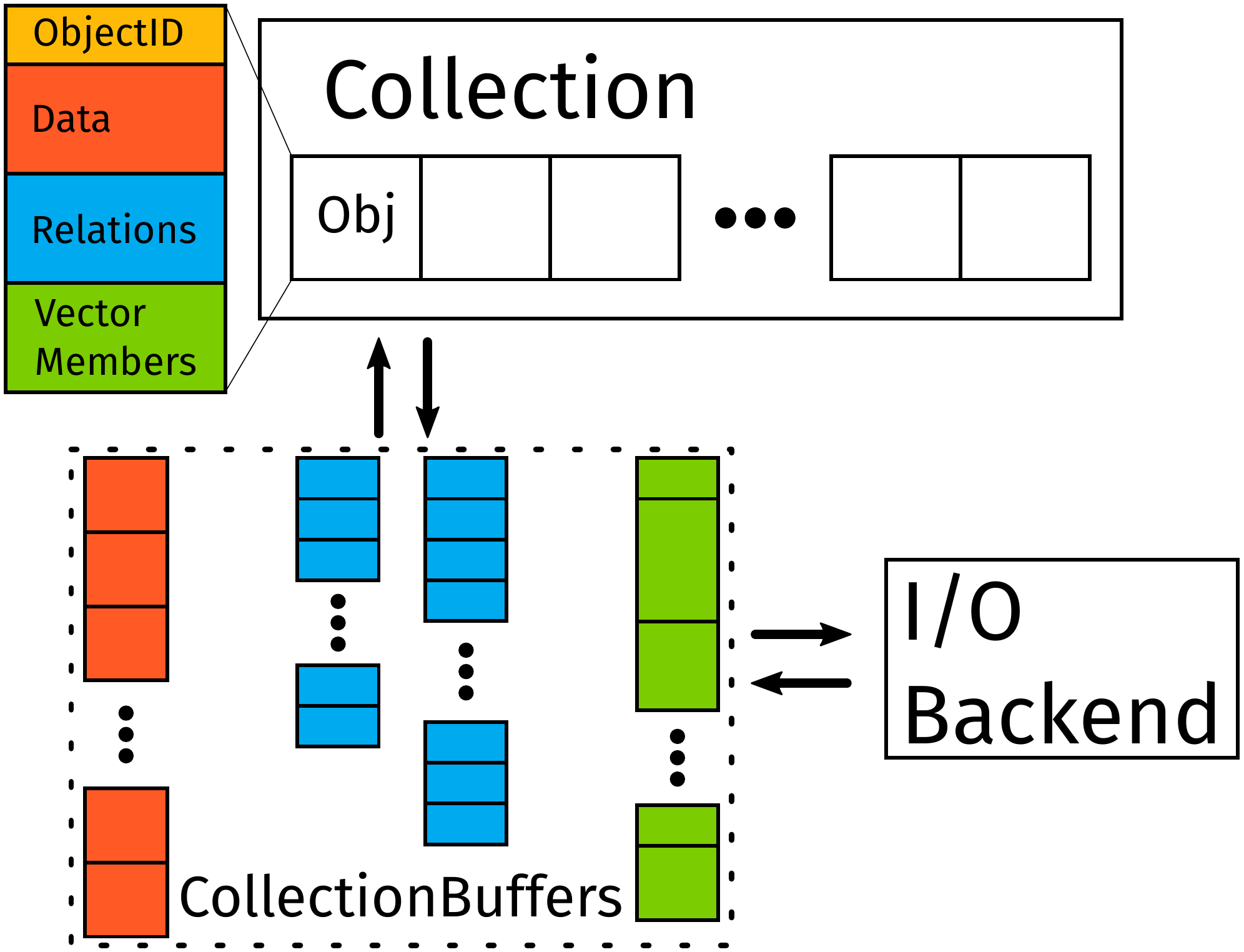}}

  \caption{(a) The three layers of podio for an example \emph{Hit} class and (b)
    the (un)packing of Collections (from) into \texttt{CollectionBuffers} for I/O.}
  \label{fig:data-layout}
\end{figure}

One of the key features of podio is its code generator that reads a high level
description of an EDM in YAML format and generates performant \cpp code which is
then compiled into a shared library. The generated code favors composition over
inheritance and uses plain-old-data (POD) types wherever possible~\cite{podio}.
The design leverages three different layers which allows for a performant I/O
layer and cache friendly memory layout. These layers are shown in
Figure~\ref{fig:podio-layers} and organize the generated classes into the
following:
\begin{itemize}
  \item The \emph{User Layer} which is the only one with which users interact,
        consisting mainly of thin handles.
  \item The \emph{Object Layer} that manages resources and inter-object
        relations.
  \item The \emph{POD} or \emph{Data Layer} where the data of all objects live
        as simple POD structs.
\end{itemize}

Each object in the Object Layer is uniquely identified by an \texttt{ObjectID}
consisting of a collection ID and an index in that collection. The collections
contain these objects and give access to the data via thin handle classes. These
handles come in two varieties; the default handles are immutable and lack any
functionality that would allow one to alter the state of the underlying object. On
the other hand the collections also serve as factories for mutable handles that
can be used to fill the collections with meaningful content. An implicit
conversion of mutable to immutable handles makes sure that interfaces consuming
EDM datatypes only need to be defined once.

Data persistency in podio generated EDMs is based on collections of objects.
Collections can be created from or grant access to their data in the form of
\texttt{CollectionBuffers}. These hold all the necessary data to (de)serialize a
collection and they are comprised solely of contiguous arrays of PODs (in
\emph{array-of-struct (AoS)} layout) as shown in Figure~\ref{fig:collection-io}.
The necessary functionality to (un)pack these buffers into the in-memory
representation of the stored objects is part of the generated EDM code. Hence,
new I/O backends effectively only need to be able to read and write contiguous
arrays of memory.

\section{The Frame concept and multithreading}\label{sec:frame}

The basic ideas of the \emph{Frame} concept have already been introduced
previously~\cite{Gaede:2022leb}. In short; the Frame is a thread-safe container that
aggregates all relevant EDM data of a given category or interval of validity,
e.g. an event or a run. Data stored in a Frame is accessible read-only and the
hand-over of the ownership of data is clearly expressed in the API. In order to
stay in line with the overall approach of offering \emph{value semantics} in the
interfaces the Frame is implemented using \emph{type erasure}. This also allows
the Frame to be constructed from almost arbitrary \emph{Frame data}, which in
turn decouples I/O operations entirely from Frame construction. As Frame data
only have to provide access to the \texttt{CollectionBuffers} once queried for them by
the Frame that owns them, they are free to defer actual work, e.g. decompressing
data, as long as possible.

This potential deferral of work is in line with the general design philosophy of
multithreading surrounding I/O of podio.
The core components of podio, e.g. readers and writers for different I/O backends, can be used to implement more sophisticated functionality in event processing frameworks. Since these usually tackle their multithreading needs in different ways, we have deliberately kept the readers and writers free of any internal synchronization and we assume that they operate on
their respective input and output files exclusively and in a single threaded
fashion. All potentially necessary synchronization, e.g. if a writer receives
multiple Frames from different threads or if a reader should supply Frame data
to a multithreaded queue, has to be done externally by the framework. For more details on how this is done for Key4hep we refer to Section~\ref{sec:frame_in_fw}.

\section{Schema evolution}\label{sec:schema_io}

Schema evolution, i.e. the possibility of an EDM schema to adapt to changing
requirements, e.g. from new detector technologies or from novel reconstruction
algorithms, is a crucial feature. It is crucial for data preservation efforts and
long term usability of any EDM. In order to avoid having to solve schema
evolution in all generality, we have chosen to implement the necessary
evolutions as they arise in EDMs that are generated by podio. This still leaves
a huge task to tackle, and was undoubtedly the most challenging development in
podio so far. The main considerations for schema evolution for podio were
\begin{itemize}
  \item being able to leverage existing capabilities of the backend, e.g. ROOT,
  \item having schema evolution work for all backends, and
  \item automating the generation of the necessary evolution code, while still
        allowing for user overrides if necessary.
\end{itemize}
As of the writing of these proceedings not all of these goals have yet been
fully achieved. Nevertheless, many of the necessary building blocks are in
place, and schema evolution is working for the default ROOT backend.

In our current design of schema evolution all the necessary evolutions are
applied to the \texttt{CollectionBuffers} directly, before collections are even
constructed from them. Hence, users will only ever see the latest version of the
EDM in memory. The evolution of the buffers will also always happen in one
evolution step, going directly from the schema version on file to the current
schema version of the EDM.

\subsection{Building blocks for schema evolution}\label{sec:schema_parts}

The first step in schema evolution is to detect changes between two EDM schemas.
To this end we have implemented a tool that reads two versions of the schema in
the high level YAML format and compares them. It checks all detected evolutions
against a pre-defined list of supported evolutions. Currently, this list very
closely follows the schema evolution capabilities of ROOT. A major advantage of
having this check very early in the process of generating code is that we can
inform users about a potentially unsupported schema evolution before they are
able to write data that is not backwards compatible with previously written
data.

The second step is to read back the \texttt{CollectionBuffers} in the schema version
they have been written. We achieve this via a central
\texttt{CollectionBufferFactory} that is able to create empty \texttt{CollectionBuffers}
for all known datatypes and schema versions. It is populated during dynamic
library loading of a generated EDM and is effectively implemented as a map of a
pair of datatype name and schema version to a buffer creation function. These
creation functions as well as the necessary call to register them into the
\texttt{CollectionBufferFactory} are all done via automatic code generation. Since the
factory is immutable during the runtime of a program it is accessible
concurrently from multiple readers that might live on several threads.

After reading the data buffers in an old schema version the final step is to
evolve these buffers to the current schema version so that a collection can be
constructed from them. We follow a very similar approach here as we did for
buffer creation by employing another central \texttt{SchemaEvolution} instance
that keeps track of evolution functions for all datatypes and schema versions.
The major difference with respect to the \texttt{CollectionBufferFactory} is that here we allow
the user to override automatically generated evolution functions if desired or
necessary.

\begin{figure}[h]
  \centering

  \includegraphics[width=0.6\linewidth]{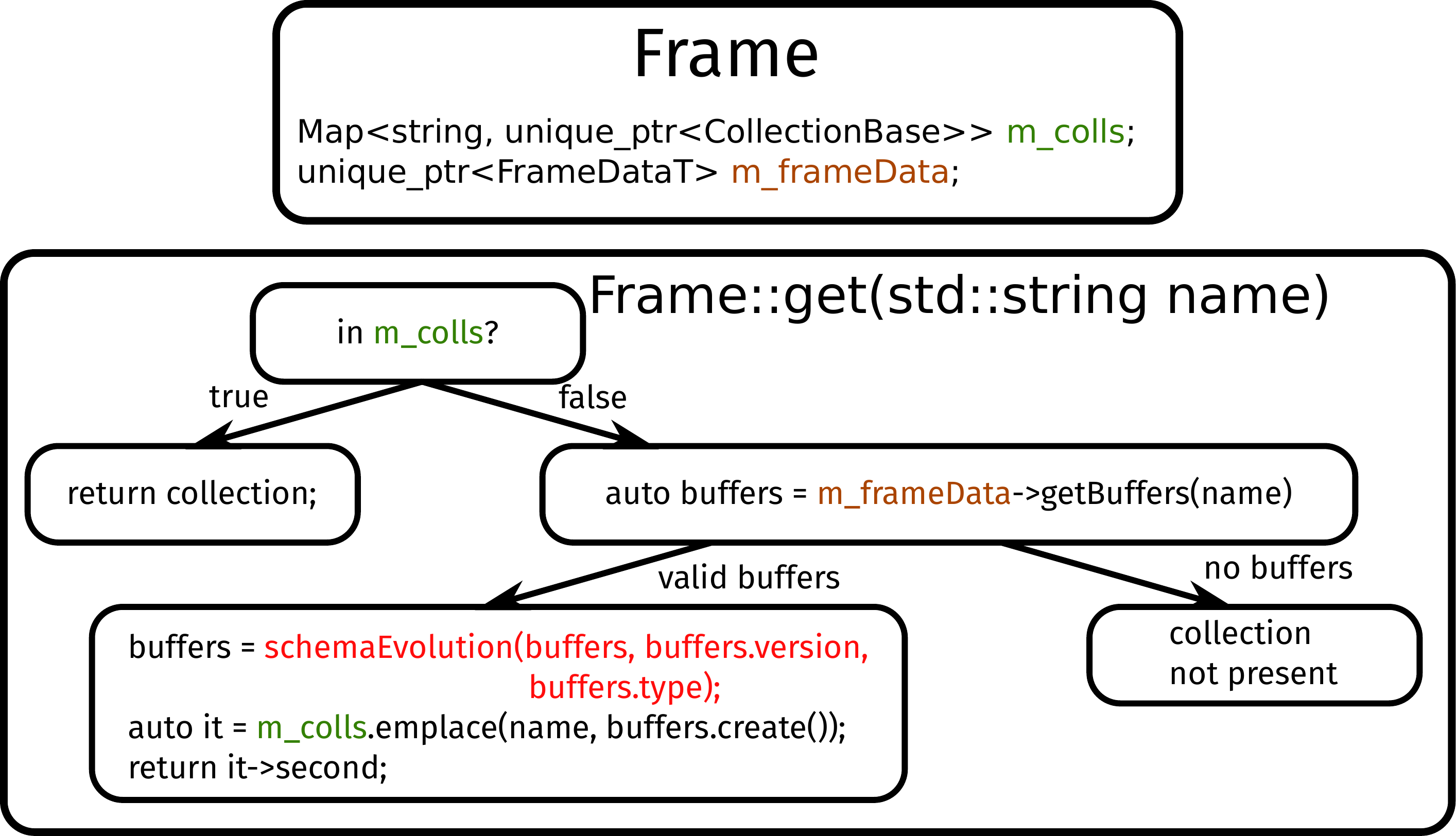}

  \caption{Minimal schematic Frame implementation and control flow inside
    \texttt{Frame::get} when requesting a collection by name.}
  \label{fig:frame_get}
\end{figure}

The hooks to execute the evolution function are placed inside the Frame directly
after obtaining the \texttt{CollectionBuffers} from the Frame data. This is the earliest
place where \texttt{CollectionBuffers} for a collection are actually guaranteed to exist,
and also the latest place where we want to deal with older schema versions. It
also makes it very easy for backends that have builtin schema evolution to
simply provide already evolved buffers in which case the schema evolution will
be a no-op. A schematic control flow with some details about the Frame internals
are shown in Figure~\ref{fig:frame_get}. As evident from there the
\texttt{CollectionBuffers} also carry enough information about the collection type and
the schema version to be able to identify the correct evolution function.

\subsection{Current status}\label{sec:schema_status}

As of the writing of these proceedings, schema evolution is working for the ROOT
backend, mainly because it's the ROOT backend who does most of the heavy lifting. The only feature
which is not natively supported is the renaming of data members, for which we
implemented the necessary evolution code. For technical reasons this evolution
path actually foregoes most of the building blocks described above. It does not
make use of the \texttt{SchemaEvolution} function registry for example, and always
requests buffers in the current schema version from the \texttt{CollectionBufferFactory}.
This is necessary to actually trigger the schema evolution mechanism of ROOT.

For the support of other backends the necessary schema evolution building blocks
are in place. This includes the code generation for populating the
\texttt{CollectionBufferFactory} with the required buffer creation functions. The main
missing piece is the code generation of the actual schema evolution functions.
However, it would in principle already be possible now to provide user defined
evolution functions to achieve schema evolution with backends without builtin
schema evolution.

We have implemented easy to re-use and automated test cases for the currently
supported schema evolution capabilities. These allow us to add tests for other
backends with very little effort.

\section{Other recent developments}\label{sec:recent_devs}

There have also been several other smaller recent developments in podio. Most of
them involved refactoring existing code to allow us to implement all the necessary
functionality for schema evolution and the Frame concept. These were completely
transparent for users. Additionally, there were some additional new features
that were added to podio, as well as some developments that were not completely
transparent. We will discuss these further in the following paragraphs.

\subsection{Stable collection IDs}\label{sec:stable_coll_ids}

For somewhat historical reasons the initial assignment of collection IDs to
collections was based on the insertion order into a Frame. This approach is
untenable in a multithreaded context as insertion order can no longer be
guaranteed. Additionally, collection IDs are a core part in the persistency of
inter-object relations. Hence, a given collection name ideally always maps to
the same collection ID, such that relations might even be reconstructed on files
that have been split during processing.

To achieve this we have switched the calculation of the collection ID to use the
\emph{MurmurHash3} algorithm to compute a 32 bit hash value from the collection
name. The choice of 32 bits was on the one hand based on the fact that the
collection ID used 32 bits before, so that this change could be implemented
effectively transparently for users and we would not need to increase the size
of the ObjectIDs by 50\%\footnote{The ObjectID consists of two 32 bit integers, a CollectionID and an ObjectIndex.} to go to 64 bits. On the other hand we also checked
that there are no collisions in the collection names that are currently in use
inside the Key4hep project. We are confident that at least for the foreseeable
future there should be no collisions in the collection IDs. However, even if
there are, a switch to 64 bits would be completely backwards compatible.

\subsection{Persistency of the datamodel definition}\label{sec:edm_def_storing}

Although podio now provides a schema evolution mechanism, sometimes it might be
useful or necessary to read the data of a file back with the original schema
version, or to simply compile an EDM library to read the data back in the first
place. In order to make it possible to retrieve the original version of the
datamodel definition in YAML format we have started to embed it into the shared
EDM library itself, but also to store it as meta data into all data files that
are written by podio. In both cases the datamodel definition is converted into the
JSON format as that allows for a more compact string representation. This string
is embedded as a raw string literal into the shared EDM library, which can be
retrieved with various tools that allow one to read binary files, e.g. via
\emph{readelf}. For the retrieval of the datamodel definition from datafiles we have
integrated the necessary functionality into the \texttt{podio-dump} utility.

\subsection{Adding an RNTuple I/O backend}\label{sec:rntuple}

\emph{RNTuple} is the designated successor of ROOTs TTree data
format~\cite{Blomer:2020usr,Lopez-Gomez:2022umr} and will address future needs
for I/O in HEP. Since podio is used for future collider studies it is natural to
strive to also support this feature as an additional I/O backend. Given the nice
separation of I/O concerns and operating on the data introduced with Frame based
I/O, there were no real technical challanges to overcome from the podio side.
The major issues that we faced had to do with trying to persist data types that
were not yet supported by RNTuple, e.g. \texttt{uint16\_t}, but these were
promptly resolved by the ROOT developers, so that the RNTuple based backend in
the end just requires a slightly more recent version of ROOT\@. Given that the
RNTuple interface is still in experimental state, also the RNTuple backend of
podio should be considered experimental and not used for production.

\subsection{Frame integration into k4FWCore}\label{sec:frame_in_fw}

One of the core motivations for developing the Frame concept was the support of
multithreaded frameworks and to facilitate the integration of podio based EDMs,
specifically EDM4hep, into the core, Gaudi~\cite{Barrand:2001ny} based, framework of Key4hep,
\emph{k4FWCore}~\cite{zenodo:k4fwcore}. Among the core services this packages
offers are the following:
\begin{itemize}
  \item \texttt{DataHandle}s that give access to podio based EDM collections
        inside an algorithm,
  \item a \texttt{PodioDataSvc} that manages the I/O of these collection as well
        as the creation of the necessary \texttt{DataHandle}s.
\end{itemize}

Both of these were previously implemented partially in terms of functionality
from standalone podio and some additional custom implementions of readers and
writers for podio data files. The integration of Frame based I/O into k4FWCore
also allowed us to clean up some of the differences in implementation that have
grown over time w.r.t. standalone podio. Here we have simply replaced the custom
readers and writers for podio files in k4FWCore with the ones that standalone
podio offers.

For users of the Gaudi based framework the Frame is not visible and is instead
just used behind the scenes to implement the \texttt{DataHandle}s and the \texttt{PodioDataSvc}.
For most existing algorithms this was a completely transparent change. However,
if an algorithm made use of the \texttt{PodioDataSvc} directly some interface changes
were unavoidable and required intervention. In order to allow for a gradual
migration to the new functionality we have kept the pre-existing functionality
in the \texttt{PodioLegacyDataSvc}. This will be deprecated and then removed in
the midterm future.

To facilitate access to file level meta data we also introduced a
\texttt{MetaDataHandle} that works similar to the \texttt{DataHandle} with the major
difference being that it can only be accessed for writing during algorithm
initialization or finalization, i.e. when Gaudi is running on a single thread.
It is backed by another Frame and makes use of the standard podio I/O
functionality. The switch to the \texttt{MetaDataHandle} had the biggest impact on
algorithms that made use of \emph{collection parameters}, e.g. encoding strings
for interpreting bit field values in the data. Although these parameters never
changed in practice, it would have still been possible to change them from a
technical perspective. Applying the now more restrictive, but thread safe,
scheme for meta data access required some work in those algorithms that need to
write meta data. A particular challenge, for which we do not yet have a fully
satisfactory solution, is algorithms for which these meta data only become
available during event processing. Here we have resorted back to requiring
execution on a single thread, which then also allows us to slightly relax the
conditions for meta data write access.

\section{Conclusion \& Outlook}\label{sec:conclusions}

The podio EDM toolkit has been and is currently used by several communities to
do physics studies. The file format that is written by podio has been stable
since the introduction of Frame based I/O, and we try to keep backwards
compatibility for these files even before a v1.0 release. Having finished a
first version of schema evolution capabilities for the default ROOT backend it
is now also possible to actually evolve datamodels that make use of podio.
Nevertheless, there are still some developments that we want to finalize before
we announce a first stable version of podio.

The most important of these developments are related to schema evolution. The
most important part that is not yet implemented here is the handling of multiple
older schema versions. We expect this to be solvable with reasonable effort
since the main work will be related to handling multiple input YAML files,
whereas all the complicated issues in code generation are already solved. We
plan to defer the support for schema evolution for non-ROOT backends until after
the release of v1.0, as this should be fairly independent and does not have an
impact on the file format written by the ROOT backend.

Another important aspect is documentation. Here we have to integrate the latest
developments and features and overall organize the existing documentation into a
coherent set of pages and text. We have setup basic automation for the
generation of documentation from the source code as well as from existing text
documents and we plan to extend this to also cover the deployment of
documentation onto a webpage.

A potential development that we plan to investigate after the release of v1.0 of
podio is related to the in-memory layout of the data in podio. Currently the
data in the Data Layer, see Figure~\ref{fig:podio-layers}, are stored in AoS
layout. However, there might be some performance gains to be made by switching
to a different layout, e.g. \emph{struct-of-array (SoA)}. This layout can be
more cache friendly or allow for better optimiziations if operations are only
necessary on a subset of members of a given data type. It would potentially also
allow for a more easy hand-off to GPUs or other heterogeneous resources. Since
podio is already generating code that features the three implementation layers,
described in Section~\ref{sec:podio_basics}, we think that we can implement this
transparently for users. However, we have not yet started any developments in
this direction.

To summarize and finish these proceedings, we want to again highlight the fact
that podio is in use by several communities in production already. The
introduction of schema evolution and the Frame concept should make podio future
proof. After having resolved the few remaining issues mentioned in the text, we
plan to release v1.0 of podio shortly and continue to incorporate the feedback
and feature requests from the user communities afterwards.

\section*{Acknowledgements}
This work beneﬁted from support by the CERN Strategic R\&D Programme on
Technologies for Future Experiments
\linkk{https://cds.cern.ch/record/2649646/}{(CERN-OPEN-2018-006)}.

This project has received funding from the European Union's Horizon 2020
Research and Innovation programme under Grant Agreement no. 101004761.

\bibliography{bibliography.bib}

\end{document}